\documentclass[namedreferences]{solarphysics}

\usepackage[hyperref,optionalrh]{spr-sola-addons}
\usepackage{graphicx}        

\usepackage{color}           

\chardef\us=`\_

\begin{document}

\begin{article}
\begin{opening}

\title{An optimized tidal-trigger model of the QBO, and some implications for the Carrington event}


\author[addressref={aff1},corref,email={F.Stefani@hzdr.de}]{\inits{F.}\fnm{F.}~\lnm{Stefani}}\sep
\author[addressref={aff1}]{\fnm{G.M.}~\lnm{Horstmann}}\sep
\author[addressref={aff1}]{\fnm{G.}~\lnm{Mamatsashvili}}\sep
\author[addressref={aff1}]{\fnm{T.}~\lnm{Weier}}\sep
\address[id=aff1]{Helmholtz-Zentrum Dresden -- Rossendorf, Bautzner Landstr. 400,
D-01328 Dresden, Germany}

\runningauthor{F. Stefani {\it et al.}}
\runningtitle{An optimized tidal-trigger model of the QBO, and 
some implications for the Carrington event}

\begin{abstract} 
Magneto-Rossby waves in the solar tachocline are currently being discussed as a potential cause of the quasi-biennial oscillation (QBO). By analyzing sequences of ground-level enhancement (GLE) events and S-flares, the dominant period of the QBO was recently shown to be close to 1.723 years, which is the dominant beat between the periods of the two-planet spring tides of Venus, Earth and Jupiter. We improve upon this model by taking into account the dependence of the three tidally-triggered magneto-Rossby waves on the actual strength of the toroidal field at the tachocline, which we 
infer from the averaged monthly sunspot number.
When optimizing the parameters of this magnetic-field 
dependence, the correlation of the tidal-forcing function with 
the 109 extreme solar events reaches values of up to 0.8.
This is much higher than the corresponding value for the field-independent tidal forcing function ($\approx 0.4$), and also higher than the correlation 
with the sunspot number ($\approx 0.56$).
Based on this improved model, we discuss some interesting parallels between the Carrington event of 1859 and the clustering of strong solar events in summer and autumn 1989. We also make some cautious forecasts for the remainder of cycle 25.

\end{abstract}
\end{opening}
\section{Introduction}

While the key role of Rossby waves in weather on Earth 
has been known for almost a century \citep{Rossby1939}, 
similar waves may also be important for solar activity 
and space weather  
\citep{Zaqarashvili2010a,Zaqarashvili2010b,Dikpati2012,Marquez2017,Dikpati2018,Gachechiladze2019,Bilenko2020,Dikpati2020,Zaqarashvili2021}. 
This applies, in particular, to the quasi-biennial oscillation (QBO)
\citep{Bazilevskaya2014}, which has attracted  considerable interest 
 since its discovery in coronal holes by \cite{McIntosh1992}, and in 
 cosmic rays and open magnetic flux by \cite{Valdez1996} and
\cite{Rouillard2004}.
As shown by \cite{Zaqarashvili2010a}, 
the interplay between differential rotation and the 
strong toroidal magnetic field at the tachocline results in the instability of magneto-Rossby waves therein with a period of 
approximately two years.
Moreover, \cite{Raphaldini2019}  argued that the dynamics of a 
resonant triad of magneto-Rossby waves could lead to periodically 
changing wave amplitudes with periods comparable to the 
dominant 11-year Schwabe cycle.

In this respect, a most interesting feature of magneto-Rossby waves is that their 
typical eigenperiods of a few hundred days 
fit remarkably well with the spring-tide periods of the tidally 
dominant planets Venus, Earth and Jupiter.
A detailed mathematical study of the resonance conditions led
\cite{Horstmann2023} to conclude that realistic-amplitude spring tides may trigger magneto-Rossby waves with velocities of the order of m s$^{-1}$ or larger, depending on a damping parameter whose precise value is still unknown. The three two-planet spring tides of Venus–Jupiter (with a period of 118 days), Earth–Jupiter (199 days) and Venus–Earth (292 days) were shown by \cite{Stefani2024} to have a beat period of 11.07 years, which corresponds remarkably well to the Schwabe cycle.
This link between magneto-Rossby waves and the Schwabe cycle may provide the 
long-sought physical argument for the 
synchronization of the solar dynamo by the weak planetary tidal forces 
as suggested and discussed in a number of previous papers
\citep{Hung2007,Abreu2012,Scafetta2012,Wilson2013,Okhlopkov2016,Stefani2016,Stefani2019,Stefani2021,Charbonneau2022,Klevs2023}.

Another, much shorter beat of the three waves, with a period of 1.723 years, was
revealed by \cite{Stefani2025}.
Motivated by the remarkable finding of 
\cite{Velasco2018} that ground-level enhancement (GLE) events exhibit 
phase stability over nearly six solar cycles, we found a very precise 
agreement 
between the observed period and the theoretical one.
More recently \citep{Stefani2026}, we applied similar statistical 
methods to 37 S-flares, 
and their merger with 72 GLE events, which all showed 
phase stability and very similar periods.

Encouraged by this result, we tested also 
various functions containing the sum of the three cosines
with the periods of the spring-tides, and their time-averages.  
Again we found significant correlations of 
around 0.4. In an initial attempt to test various weights of 
the three functions, this value increased to around 0.45.

In the present paper, we aim at optimizing this correlation by 
taking the magnetic-field dependence of the excitation efficiency 
of the 
three magneto-Rossby waves into serious consideration.
To achieve this, we use the averaged monthly sunspot number as a proxy for the toroidal field at the tachocline.
As will be demonstrated, optimized parametrizations of the weights of the three waves provide correlations of
up to 0.8, which is statistically highly significant.
In this context, we will observe that a number of extreme solar events, 
which occur far from the maxima of the respective solar cycle, correspond 
nicely to the peaks of the optimized tidal-trigger function.

We will then take a look at the famous Carrington event 
of September 1$^{st}$ 1859 and ask how it may be  related to the previously 
optimized tidal-trigger function and the evolution of the 
sunspot number during solar cycle 10.
As we will see, the behaviour before and around
the Carrington event is remarkably similar to that of cycle 22, 
which exhibited a massive clustering of extreme solar events 
in the summer and autumn of 1989.

We will also make a cautious attempt to forecast on
solar activity for the remainder of cycle 25. 

The paper will conclude with a summary of the results and an
outlook on future work.

\section{Used data and employed methods}

In this section, we will describe the data utilized in this paper and discuss the methods employed to analyze them.

\begin{figure}[t]
  \centering
  \includegraphics[width=0.99\textwidth]{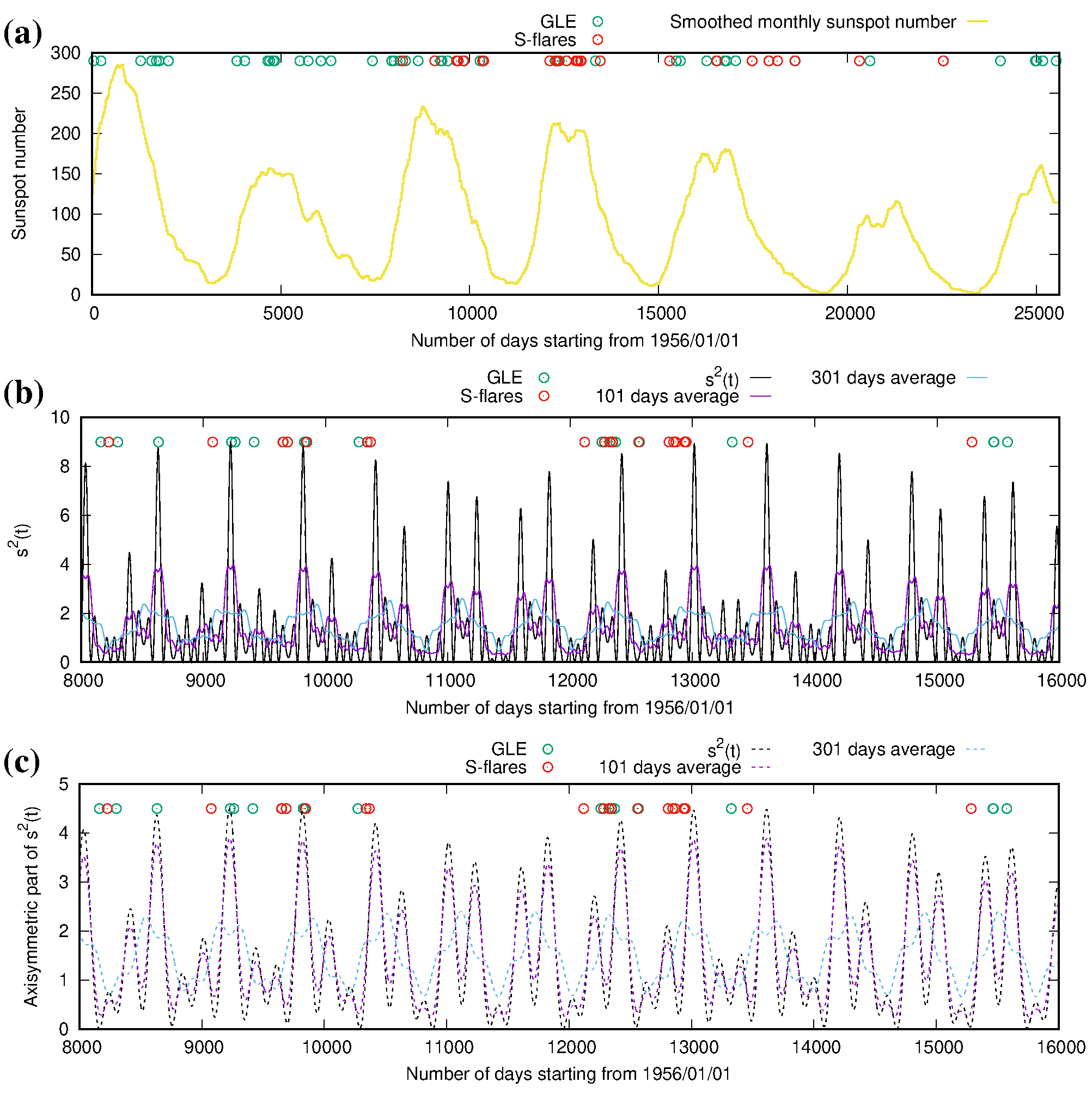}
  \caption{The 109 extreme solar events, including 72 GLE events and 37 
  S-flare events, shown alongside different
  functions. (a) Including the 13-months-averaged monthly sunspot number
  $\rm SSN$
  for the entire time interval between 1956 and 2025.
  (b) Including the function $s^2(t)$ and two representative 
  time averages of it. 
  To improve visibility, we have selected only the time segment 
  between days 8,000 and 16,000 after 1955/12/31. (c) 
  Including the axisymmetric part 
  of $s^2(t)$, together with same time-averages as in (b).}
  \label{fig:3}
\end{figure}

We start with the 13-month average of the monthly sunspot number 
as obtained from www.sidc.be/SILSO/infosnmstot .
This smoothed dataset, shown as the yellow curve in Figure 1a, is considered 
most appropriate for our purposes, since it appears to be a good proxy
for the field at the tachocline during (or shortly before) 
the extreme solar events to be considered.
As for the latter, Figure 1a shows the total of 109 extreme solar events
in the time interval between from 1956 to 2025. These
include, first, the 72 GLE-events (green), 
obtained from gle.oulu.fi, and, second, the 37 events of solar 
superflares of S-class ($>$X10 in soft X-rays) 
taken from \cite{Tan2025} and \cite{Velasco2026}. 
For the precise dates of these events, 
please refer to Table 1 (GLE) and Table 2 (S-flare) 
of  \cite{Stefani2026}.
Note that that we do not try to provide additional details
about the strength of these events. Instead, they are all 
given the same weight of 1.

Next, we will examine the tidally-triggered magneto-Rossby waves
in the tachocline.
The function
\begin{eqnarray}
s(t)&=& A \cos\left( 4\pi  \frac{t-t_{\rm VJ}}{  P_{\rm VJ}}\right) + B\cos\left( 4\pi  \frac{t-t_{\rm EJ}}{ P_{\rm EJ}} \right)+C\cos\left( 4\pi   \frac{t-t_{\rm VE}}{ P_{\rm VE}} \right) \;,
\end{eqnarray}
whose square $s^2(t)$ is shown in Figure 1b, is the sum of three cosines that
represent the Rossby waves as excited by the planetary tides. 
Their periods correspond to the two-planet spring-tides of Venus-Jupiter, Earth-Jupiter and Venus-Earth.
The specific values used here are the
synodic periods $P_{\rm VJ}=0.64884$\,years, 
$P_{\rm EJ}=1.09207$\,years,
$P_{\rm VE}=1.59876$\,years, and the 
epochs of the corresponding conjunctions
$t_{\rm VJ}=2002.34$,
$t_{\rm EJ}=2003.09$, and
$t_{\rm VE}=2002.83$, all taken 
from 
\cite{Scafetta2022}.

While the three weights $A$, $B$ and $C$ will play a key role in the following sections, for generating the plots in Figure 1 they were all set to 
unity.  In addition to the function $s^2(t)$, we also show 
two representative time averages with 
moving-average windows  of 101 and 301 days.

The full-line curves shown in Figure 1b refer to 
zero azimuthal angle, while the dashed curves in Figure 1c show the 
corresponding  averages over the azimuth, i.e. the axisymmetric 
part of $s^2(t)$. Note that
it is by no means a given that the curves in 
Figure 1b and 1c are so similar.
Actually, this similarity has to do with the notion of 
``orbital invariance'', 
defined by \cite{Scafetta2022}.

Even a mere visual inspection reveals  a certain
degree of synchronizm between the solar events and 
the averaged sunspot number on one hand 
(Figure 1a), and the peaks of the function $s^2(t)$
(and its time and/or azimuthal averages) on the other hand
(Figures 1b,c).

To quantify this connection, we compute  
the correlation 
\begin{eqnarray}
{\rm Corr}&=&\frac{ \sum_{i=1}^{N} [f(t_i)-\overline{f(t)}]   }
{\sqrt{\sum_{i=1}^{N}  [f(t_i)-\overline{f(t)}]^2}} \; ,
\end{eqnarray}
between 
the $N=109$ extreme solar events with a 
given function $f(t)$, as introduced in \cite{Stefani2026}.
If we use for $f(t)$ the averaged monthly sunspot number, $\rm SSN(t)$, we obtain a correlation of 0.56. Such a  high value is not surprising, given that extreme solar events often, although not always, occur around the maximum of the solar cycle.

What is more remarkable is the fact that the tidal-trigger function $s^2(t)$  
also exhibits statistically significant correlations with solar events. 
In Table 1 we provide
the corresponding values of $\rm Corr$
for different time and/or azimuthal averages.

\begin{table}[]
\scriptsize
\caption{Correlations $\rm Corr$ of $s^2(t)$, its axisymmetric part, and various time 
averages with the 109 extreme solar events for the simple 
equal-weight parameter choice $A=B=C=1$. The maximum correlation 
values in the  third column occur at the optimal time-shift between 
the function and the solar events that is indicated in the second column.
Here, a positive value means that the highest correlation
occurs for $s^2(t)$ taken at later times than the 
very solar events.}
\begin{tabular}{cccc}
\hline
\hline
 & Average (days) & Optimum shift (days)    & $\rm Corr$ \\
\hline
$s^2(t)$ & 1 & 80 & 0.335 \\
&61 &74 & 0.339 \\
&101 &51 & 0.307 \\
& 201 & 12 & 0.396 \\
& 301 & 65 & 0.380 \\
& 401 & 21 & 0.463 \\
Axisymmetric $s^2(t)$ & 1 & 94 & 0.276 \\
& 61 & 92 & 0.286 \\
& 101 & 89 & 0.302 \\
& 201 & 33 & 0.370 \\
& 301 & 15 & 0.414 \\
& 401 & 15 & 0.430 \\
\hline
\end{tabular}
\end{table}

The discussion up to this point corresponds basically to the
state of affairs as described in \cite{Stefani2026}.
The obtained correlations of around 0.4 are already highly 
significant, with a $p$-value around $10^{-5}$ for the considered 
109 events.
The question now is whether this correlation can 
be further enhanced with an improved 
physical modeling.

To do this, we will take a closer look at the weights $A$, $B$ and $C$ in Equation 1. On the one hand, they are proportional to the respective spring-tide triggers of the planets. Of these, the Venus-Jupiter combination (118 days) is stronger than the other two.
On the other hand, as shown in Figures 2, 3 and 4 of \cite{Stefani2024}, 
the amplitude of tidally-triggered Rossby waves depends 
also, and in  a
non-trivial manner, on the toroidal field 
$B_{\varphi}$ prevailing 
at the tachocline.  Roughly speaking, the wave amplitude is quite small for vanishing fields and generally increases as the field increases. 
However, as exemplified in Figure 4 of \cite{Stefani2024}  for the wave 
with $P_{VE} = 292$\,days, it can also happen 
that the tidal excitation completely is completely suppressed 
in the case of a 
field that is too strong, exceeding (for the specific parameters of 
Figure 4) some 40\,kG (i.e., 4\,T).

The first step in the physical modeling of the toroidal field effect 
is to infer its value $B_{\varphi}$ at the tachocline level from the
averaged sunspot number. We use here the 
relation
\begin{eqnarray}
B_{\varphi}=5.0 \sqrt{{\rm SSN}/141.7} \; {\rm T} \;,
\end{eqnarray}
which is motivated by applying a typical dynamo-scaling relation to
the average sunspot number 141.7 measured in 1956, and assuming
a field at the tachocline of 5\,T for that time.
Note, however, that the precise numerical factor is 
unimportant as it will be absorbed in an overall scaling 
factor when carrying out the optimization process.

In a second step, we parametrize the field dependence of the weights in Equation (1) in the  
form
\begin{eqnarray}
A(B_{\varphi})&=&(a_0+a_2 B^2_{\varphi} (1-B^2_{\varphi}/a^2_4))  \nonumber \\
&&\times 0.5 (1+\rm{sgn}(a_0+a_2 B^2_{\varphi} (1-B^2_{\varphi}/a^2_4)))\;, 
\end{eqnarray}
and correspondingly for $B(B_{\varphi})$ and $C(B_{\varphi})$. 

The parameters in Equation 4 can be interpreted as follows:
The value of $a_0$ (and, by extension, $b_0$ and $c_0$) is a measure
for the excitation efficiency of the Rossby waves in the absence of
any prevailing magnetic field $B_{\varphi}$. 
The value of $a_2$ (as well as $b_2$ and $c_2$) indicates the 
general sensitivity of wave excitation on the magnetic field 
strength. Finally, $a_4$ (as well as $b_4$ and $c_4$) has the dimension 
of a magnetic field and determines the field strength at which the wave excitation breaks down. Therefore, it should be in the range of 
a few Tesla. Note that the second line of Equation (4) serves 
just to avoid negative values.

Aiming at maximizing correlations, the ansatz of Equation 4 
leaves  us with nine parameters, one of which could be set to unity 
since $\rm Corr$ does not depend on an overall scaling factor of $s^2$.
Nevertheless, even the remaining eight-dimensional optimization problem 
is numerically quite expensive. In the following section, we will 
therefore test some restricted versions of it.

\section{Optimization}

In this section, we will optimize the correlation according to Equation 2 
when $f$ is identified with $s^2(t)$, or
its azimuthal and/or time-averages. 
The optimization itself is carried out using a simple 
look-up table. 
This means that the correlation is computed for each considered parameter set 
$a_i$, $b_i$ and $c_i$, and the set yielding the 
highest value of $\rm Corr$ is selected as optimum. In an intermediate step, 
we determine at 
which time-shift between the tidal-trigger function and 
the solar events $\rm Corr$ is maximum.

Table 2 and Figure 2 illustrate a simple initial example, for which 
we set $a_0=b_0=c_0=0$ and $c_2=1$ beforehand, and then vary the remaining five 
parameters, resulting in the optimized set: 
$a_2=0.7$, $b_2=0.7$, $a_4=7.6$, $b_4=4.0$, and $c_4=4.8$.

Remarkably, for the longer moving average windows 
the obtained correlation reaches values of up to 0.75, 
which is already higher than the correlation with sunspot 
numbers alone (0.56). Figure 2 shows that, as $\rm SSN$ 
approaches zero, the tidal-trigger function $s^2(t)$ drops also 
to zero. Obviously, this is because of the choice 
$a_0=b_0=c_0=0$. By contrast, on the high-field end, 
we observe a certain flattening due to the combined 
effect of the suppressed excitation and the time average.

\begin{figure}[t]
  \centering
  \includegraphics[width=0.99\textwidth]{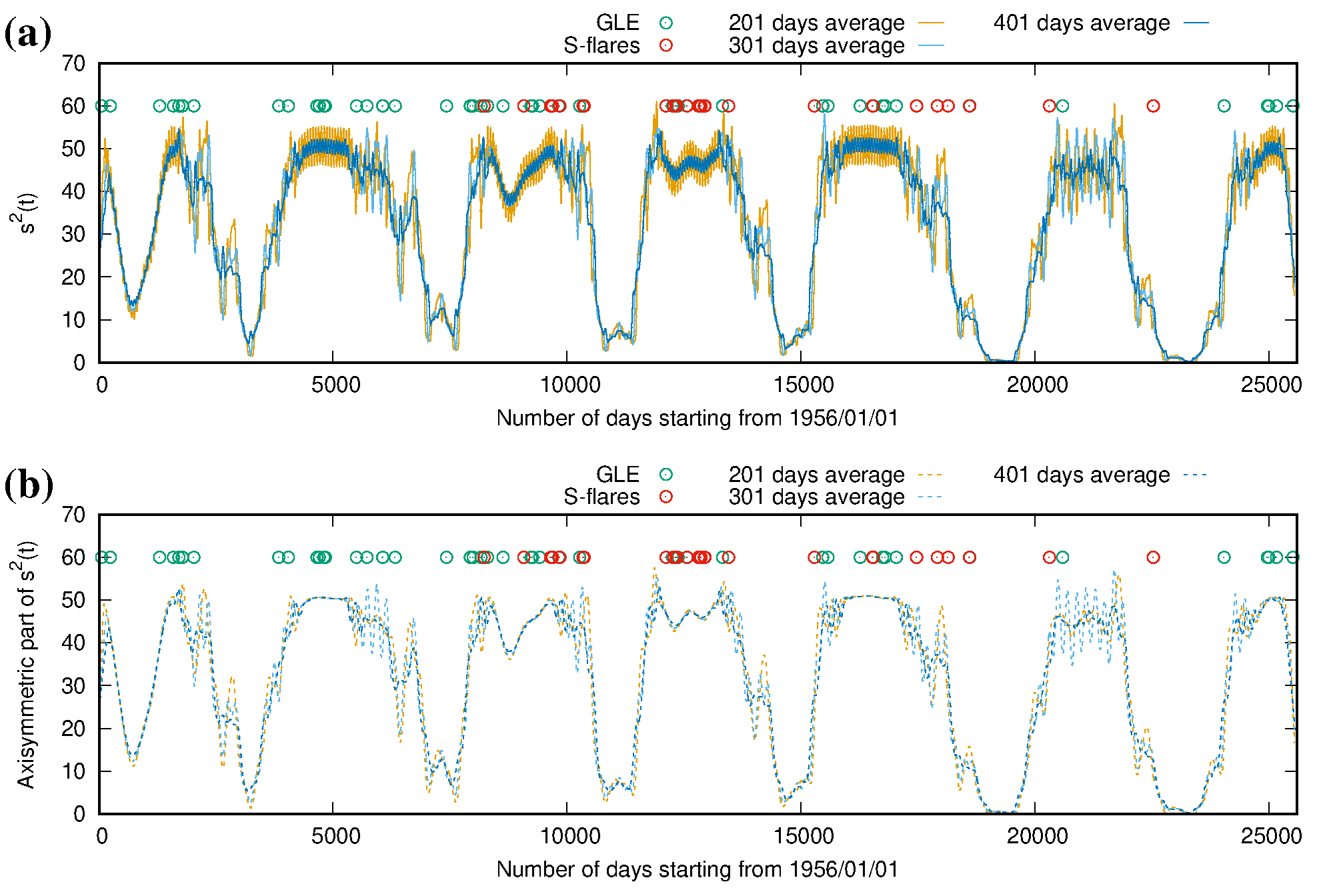}
  \caption{The 109 extreme solar events jointly with different
  averages of $s^2(t)$ for the parameter choice:
  $a_0=b_0=c_0=0$, $a_2=0.7$, $b_2=0.7$, $c_2=1$, $a_4=7.6$, $b_4=4.0$, $c_4=4.8$.
  (a) With three representative time-averages of $s^2$. (b) 
  With three representative time-averages of the 
  axisymmetric part of $s^2$.}
  \label{fig:3}
\end{figure}

\begin{table}[]
\scriptsize
\caption{Same as Table 1, but for the first optimized parameter 
set with 
$a_0=b_0=c_0=0$, $a_2=0.7$, $b_2=0.7$, $c_2=1$, $a_4=7.6$, $b_4=4.0$, $c_4=4.8$.
}
\begin{tabular}{cccc}
\hline
\hline
 & Average (days) & Optimum shift (days)    & $\rm Corr$ \\
\hline
$s^2(t)$ & 1 & 95 & 0.447 \\
&61 &118 & 0.587 \\
&101 &122 & 0.633 \\
& 201 & 20 & 0.737 \\
& 301 & -24 & 0.751 \\
& 401 & -9 & 0.744 \\
Axisymmetric $s^2(t)$ & 1 & 132 & 0.585 \\
& 61 & 124 & 0.611 \\
& 101 & 122 & 0.652 \\
& 201 & 33 & 0.724 \\
& 301 & 4 & 0.745 \\
& 401 & -15 & 0.750 \\
\hline
\end{tabular}
\end{table}

Figure 3 and Table 3 illustrate another example 
in which we have set beforehand $a_2=b_2=c_2=1$
and optimized then the remaining parameters, resulting in:
$a_0=0$, $b_0=3.0$, $c_0=5.0$,  $a_4=7.9$, $b_4=3.5$, $c_4=4.5$. 
Here, only $a_0$ 
has turned out to be zero, while 
$b_0$ and $c_0$ acquire some finite values.
This is quite plausible 
in view of the different low-field limits 
of the three waves as
shown in
Figures 2, 3, and 4 of \cite{Stefani2024}.

The correlations now reach values close to 0.8, which is 
indeed a remarkably high value.
The improvement is part due to the 
fact that $s^2(t)$ can now take on reasonably large values 
even when $\rm SSN$ is comparably small.

\begin{figure}[t]
  \centering
  \includegraphics[width=0.99\textwidth]{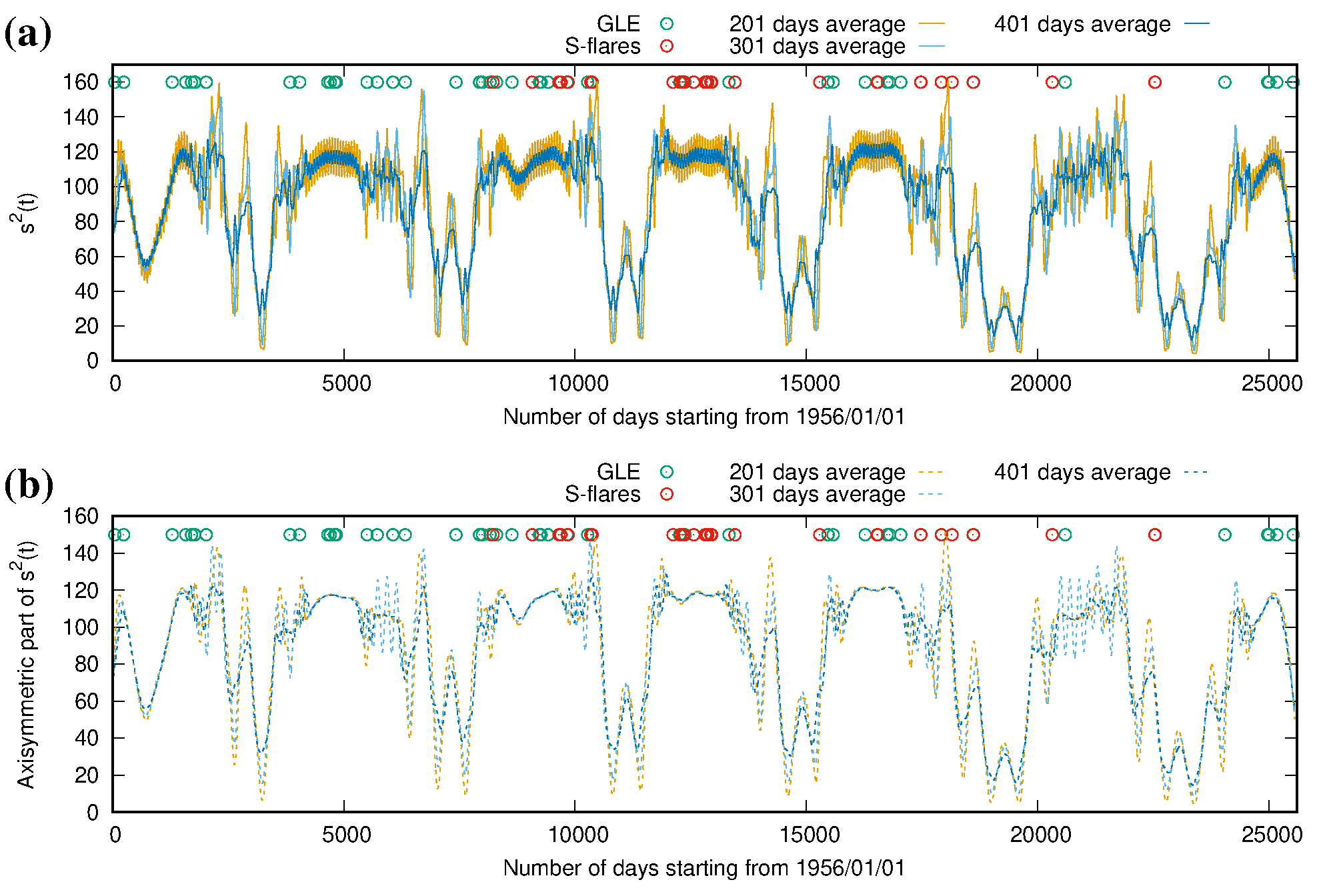}
  \caption{Same as Figure 2, but for the parameter choice:
  $a_0=0$, $b_0=3.0$, $c_0=5.0$, $a_2=b_2=c_2=1$, $a_4=7.9$, $b_4=3.5$, $c_4=4.5$.}
  \label{fig:3}
\end{figure}

\begin{table}[]
\scriptsize
\caption{Same as Table 2, but for the parameter choice:
$a_0=0$, $b_0=3.0$, $c_0=5.0$, $a_2=b_2=c_2=1$, $a_4=7.9$, $b_4=3.5$, $c_4=4.5$.}
\begin{tabular}{cccc}
\hline
\hline
 & Average (days) & Optimum shift (days)    & Correlation \\
\hline
$s^2(t)$ & 1 & 96 & 0.402 \\
&61 &115 & 0.523 \\
&101 &126 & 0.570 \\
& 201 & 22 & 0.782 \\
& 301 & -22 & 0.783 \\
& 401 & -1 & 0.785 \\
Axisymmetric $s^2(t)$ & 1 & 136 & 0.512 \\
& 61 & 130 & 0.537 \\
& 101 & 129 & 0.582 \\
& 201 & 27 & 0.746 \\
& 301 & 7 & 0.776 \\
& 401 & -19 & 0.798 \\
\hline
\end{tabular}
\end{table}

To better understand this effect, Figure 4 focuses 
on those peaks of the time-averaged $s^2(t)$-function close to 
extreme solar events of cycles 21 and 23
where $\rm SSN$ is quite small.
Some correspondences between those solar events 
and the peaks of $s^2(t)$ are indicated by magenta arrows.

\begin{figure}[t]
  \centering
  \includegraphics[width=0.99\textwidth]{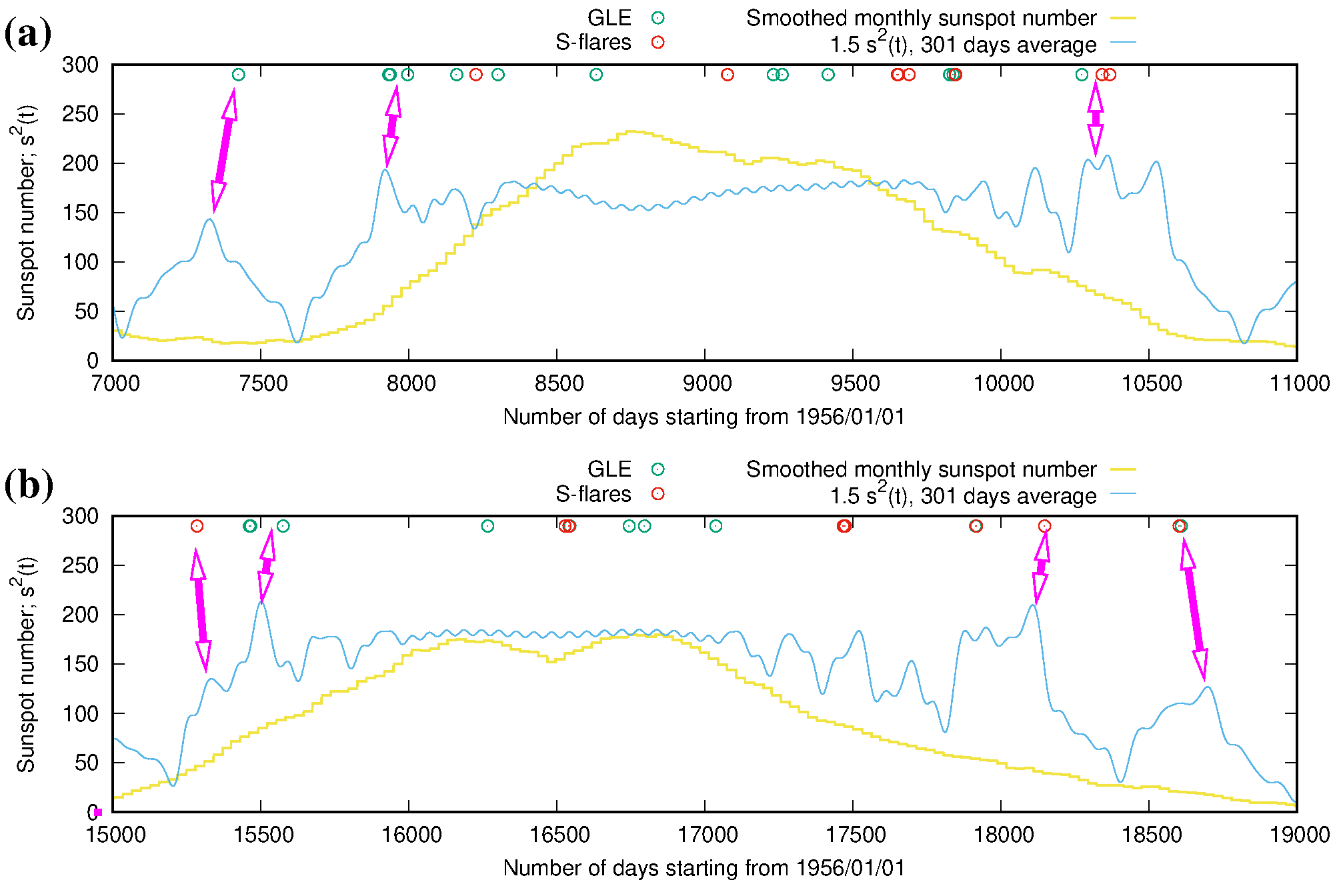}
  \caption{Zoomed-in version of Figure 3, jointly with the sunspot number 
  $\rm SSN$,
  covering solar cycles 21 (a) and 23 (b).
  The magenta arrows link peaks of the time-averaged $s^2(t)$ function
  with extreme solar events at those times where $\rm SSN$
  is comparably small.}
  \label{fig:3}
\end{figure}

\section{Carrington event}
Encouraged by the highly significant correlations 
between the optimized tidal-trigger functions 
and the 
109 extreme solar events, this section 
examines how the Carrington event \citep{Cliver2022} 
of September 1$^{st}$ 1859 
would fit into this scheme. 

This event, shown as the green open circle
in Figure 5a, corresponds to day 35,185 before
January 1$^{st}$ 1956. 
Again the yellow curve in this figure 
shows the averaged monthly sunspot number $\rm SSN$, while 
the light-blue curve show the  301-day averaged $s^2(t)$.
Over a 600-day period before the Carrington event, 
during which $\rm SSN$ steadily increased, $s^2(t)$ remained 
relatively constant. 
One might get the impression that this constancy 
enabled the Rossby waves to reach high amplitudes, 
providing then the ``final push`'' needed for the 
flux tube to launch when $B_\varphi$ just reaches its maximum.

Remarkably, a similar pattern shows up when we compare this 
behavior with that of cycle 22 (Figure 5b). Here, a large 
cluster of extreme solar events occurred during the summer and 
autumn of 1989, which seem to have a similar ``prehistory'' as 
the Carrington event of 1859. This putative link is indicated 
by the magenta dashed line.

\begin{figure}[t]
  \centering
  \includegraphics[width=0.99\textwidth]{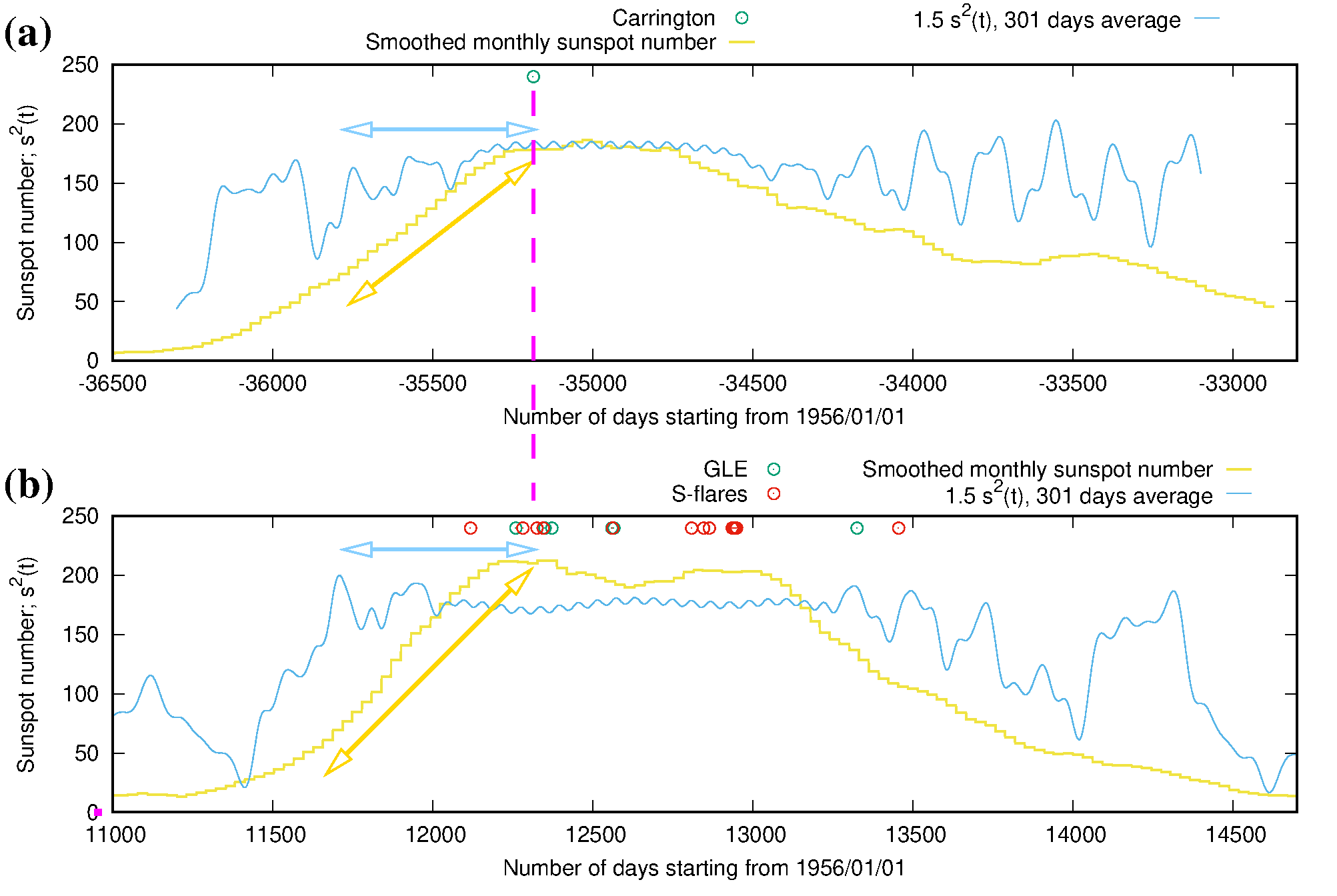}
  \caption{The Carrington event in the context of cycle 10, and 
  compared also with cycle  22. (a) Sunspot number and computed 301-day average of $s^2(t)$ 
  for cycle 10,
  with the Carrington event of 1859/09/01 indicated.
  During the 600 days before the event, $\rm SSN$ steadily 
  rose (symbolized by the yellow arrow) while 
  $s^2(t)$ remained rather constant (light-blue arrow).
  (b) The corresponding plot for cycle 22, with the clustering 
  of extreme solar events in
  summer and autumn 1989 indicated. The dashed 
  magenta line shows the putative link.}
  \label{fig:3}
\end{figure}

\section{Predictions for the remainder of cycle 25}

In this section, we ask what might be expected from our 
results for the rest of solar cycle 25.
As with any forecast, this should be considered 
with more than one grain of salt.
Evidently, the prediction of the tidal-trigger function $s^2(t)$ depends 
strongly on the accuracy of the prediction of the toroidal 
field $B_\varphi$.
For the latter, we use again Equation 3  and 
the prediction of the monthly sunspot number provided by 
NOAA under 
www.swpc.noaa.gov/products/predicted-sunspot-number-and-radio-flux.
The latter data are  shown as 
the dashed section of the yellow curve in Figure 6.
Based on this, we calculate $s^2(t)$, the 301-day time 
average of which is shown as the light-blue curve in 
Figure 6.

Obviously, after the long plateau between days 
24,600 and 25,300, a couple of peaks are predicted at 
the indicated dates.
While the first one, December 5$^{th}$ 2025, could indeed be 
related to the GLE event of November 2025 and the 
strong activity in January and February 2026, the 
subsequent dates are, of course, highly uncertain. 
Nevertheless, when the predicted curve is compared 
with those of cycles 21 and 23 (see Figure 4), it 
seems indeed possible that a few major events may still 
occur in the remainder of cycle 25.

\begin{figure}[t]
  \centering
  \includegraphics[width=0.99\textwidth]{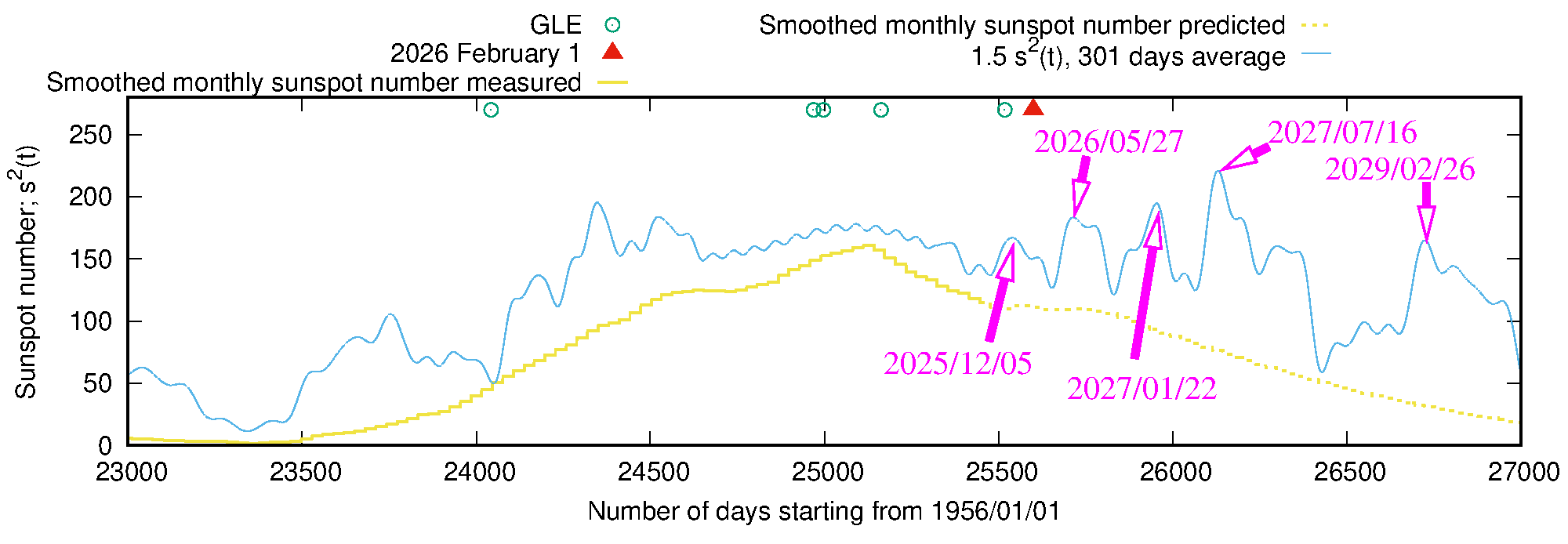}
  \caption{Predictions for the remainder of cycle 25.
  The yellow curve is divided into the measured $\rm SSN$ (full line) and 
  the predicted one (dashed line) taken from 
  ww.swpc.noaa.gov/products/predicted-sunspot-number-and-radio-flux.
  Obviously, the predicted $s^2(t)$ shows quite a couple of 
  candidate peaks in the remainder of cycle 25 at which 
  strong solar activity may be likely.}
  \label{fig:3}
\end{figure}

\section{Summary and Conclusions}

In this paper, we have continued our efforts
to explain the QBO, and the related origin of 
extreme solar events, by 
tidally-triggered magneto-Rossby waves. 
Once again, we set out from the 
hypothesis that the two-planet spring-tides of
Venus-Jupiter, Earth-Jupiter, and Venus-Earth 
trigger Rossby waves at the solar tachocline 
with periods of 118, 199 
and 292 days, respectively. While the resulting
beat of 629.29 days (1.723 years) was previously shown to be 
the dominant period underlying extreme solar events, 
the main focus laid now on the specific field 
dependence of the three individual waves.
The parametrization ansatz chosen by us was inspired by the findings 
of \cite{Horstmann2023} and \cite{Stefani2024} 
that the amplitudes of the waves typically 
increase with the toroidal field, while their 
resonant excitation can also break down in 
the presence of too strong a field.

Our preliminary attempts to optimize the parameters 
of this field dependence revealed correlations 
with the extreme solar events of up 
to 0.8, which is significantly higher than the 
respective correlation based on 
the sunspot number alone.
This suggests that the three tidally-triggered 
magneto-Rossby waves are indeed the dominant 
factor governing the occurrence of extreme solar 
events, although the specific field dependence 
of their respective weights
also plays a significant role.
It goes without saying that the parameterization and 
optimization of the field dependencies can still be
improved. Our work is just the first step in this 
direction.

A generic feature of this model is that large values of $s^2(t)$ 
may show up at relatively low values of the 
averaged sunspot numbers, which could explain  why extreme solar events can also occur at times far outside the maximum of solar cycles.

Encouraged by these results, we looked back to the 
time of the Carrington event. Relying on the 
deterministic influence of the planets, and the well-known 
sunspot number during solar cycle 10, we computed the tidal-trigger function $s^2(t)$ for that time.
We found that, over a 600-day period before the Carrington event, $s^2(t)$ formed a relatively constant plateau while the sunspot number (and hence the toroidal field) increased steadily.
This suggests that the Rossby waves had sufficient time to 
build up to strong amplitudes, while $B_\varphi$ was still 
too weak to launch flux tubes.
Then, at the moment when $B_\varphi$ reached its maximum value, 
the well-developed Rossby wave gave it the ``final kick'' needed to produce a strong rising flux tube. 
Interestingly, the comparison with cycle 22 (Figure 5b) showed a remarkably similar pattern, with a significant cluster of extreme solar events occurring during the summer and autumn of 1989.
It seems worthwhile to investigate this  
interaction between the slowly growing $B_\varphi$ and the steady accumulation of energy in the Rossby wave in more detail.

Ultimately, we ventured to predict the level of 
solar activity for the remainder of cycle 25.
Based on a forecast of the steadily decaying sunspot 
number, we identified a number of candidate peaks
of $s^2(t)$ at which the solar activity 
might be prone to further strong events.
Although such events would not be surprising given 
similar occurrences during the decline phases of cycles 
21 and 23, forecasts of this kind should certainly be treated 
with great caution.

While the present work focused on explaining extreme solar events in terms of the 
tidal-trigger function and its dependence
on the {\it observed} magnetic field, future work should aim to 
compute this very field in a self-consistent manner.
Actually, all dynamo simulations
carried out previously \citep{Stefani2019,Stefani2020,Stefani2021,Klevs2023,Stefani2024,Stefani2025} 
were mainly dedicated to the understanding 
of the ``planetary clock''  of the Schwabe cycle, and 
of longer time cycles such as Suess-de Vries and 
Gleissberg.
In a sense, the current focus on explaining 
the QBO and extreme solar events has taken us a 
step back from this final goal. However, it has 
not been forgotten.

\begin{acks}
F.S. would  like to thank  
Tony Phillips for inquiring into potential impacts 
of the January 2026 alignment on solar activity, and
Lakshmi Pradeep Chitta for an inspiring discussion about 
the ``postdictability'' of the Carrington event.
\end{acks}
\\
\\

\begin{footnotesize}
\noindent
{\bf Funding information} This work received
funding from the Helmholtz Association 
in frame of the AI project GEOMAGFOR (ZT-I-PF-5-200),
and from Deutsche Forschungsgemeinschaft (DFG) under Grant No. MA10950/1-1.
\end{footnotesize}

\section*{Disclosure of Potential Conflicts of Interest}
The authors declare that they have no conflicts of interest.

\end{article} 

\end{document}